\documentclass[conference]{IEEEtran}
\IEEEoverridecommandlockouts
\usepackage{cite}
\usepackage{amsmath,amssymb,amsfonts}
\usepackage{graphicx}
\usepackage{textcomp}
\usepackage{xcolor}
\def\BibTeX{{\rm B\kern-.05em{\sc i\kern-.025em b}\kern-.08em
    T\kern-.1667em\lower.7ex\hbox{E}\kern-.125emX}}
    
\usepackage{amsmath}
\usepackage{amsthm}
\usepackage{amssymb}
\usepackage{csquotes}
\usepackage[normalem]{ulem}

\usepackage{hyperref}

\theoremstyle{plain}

\theoremstyle{definition}

\usepackage{listings}
\usepackage{longtable}
\usepackage{booktabs}
\usepackage{enumitem}
\usepackage{xcolor}

\definecolor{RO2Blue}{HTML}{1D4ED8}   
\definecolor{RO2Teal}{HTML}{0F766E}   
\definecolor{RO2Green}{HTML}{16A34A}  
\definecolor{RO2Frame}{gray}{0.80}    

\lstdefinestyle{RO2style}{
  language=Python,
  basicstyle=\ttfamily\small,
  keywordstyle=\color{RO2Blue}\bfseries,
  commentstyle=\color{RO2Teal},
  stringstyle=\color{RO2Green},
  numbers=left,
  numbersep=6pt,
  stepnumber=1,
  frame=single,
  rulecolor=\color{RO2Frame},
  breaklines=true,
  tabsize=2,
  showstringspaces=false,
  upquote=true,
  columns=fullflexible
}
\usepackage{siunitx}
\DeclareSIUnit{\uw}{\micro\watt}
\DeclareSIUnit{\uj}{\micro\joule}
\DeclareSIUnit{\TWh}{\tera\watt\hour}

\usepackage{graphicx, color}
\graphicspath{{fig/}}

\usepackage{algorithm, algpseudocode} 
\usepackage{mathrsfs} 

\hypersetup{
colorlinks=true,
linkcolor=black,
citecolor=black,
anchorcolor=black,
filecolor=black,
urlcolor=black
}

\usepackage{caption}
\usepackage{subcaption}
\usepackage{pdfpages}
\usepackage{comment}

\usepackage{subcaption} 

\usepackage{mdframed}

\newcommand{\code}[1]{\begin{footnotesize}\textsf{#1}\end{footnotesize}}

\usepackage{amsmath}
\makeatletter
\newcommand{\xleftrightarrow}[2][]{\ext@arrow 3359\leftrightarrowfill@{#1}{#2}}
\newcommand{\xdashrightarrow}[2][]{\ext@arrow 0359\rightarrowfill@@{#1}{#2}}
\newcommand{\xdashleftarrow}[2][]{\ext@arrow 3095\leftarrowfill@@{#1}{#2}}
\newcommand{\xdashleftrightarrow}[2][]{\ext@arrow 3359\leftrightarrowfill@@{#1}{#2}}
\def\rightarrowfill@@{\arrowfill@@\relax\relbar\rightarrow}
\def\leftarrowfill@@{\arrowfill@@\leftarrow\relbar\relax}
\def\leftrightarrowfill@@{\arrowfill@@\leftarrow\relbar\rightarrow}
\def\arrowfill@@#1#2#3#4{%
  $\m@th\thickmuskip0mu\medmuskip\thickmuskip\thinmuskip\thickmuskip
   \relax#4#1
   \xleaders\hbox{$#4#2$}\hfill
   #3$%
}
\makeatother

\begin{document}

\title{Spanergy: Energy-aware Distributed Tracing for Microservices}

\author{\IEEEauthorblockN{César Perdigão Batista, Denis Conan, Sophie Chabridon}
\IEEEauthorblockA{SAMOVAR, Télécom SudParis, Institut Polytechnique de Paris, 91120 Palaiseau, France\\
\{cesar-augusto.perdigao\_batista, Denis.Conan, Sophie.Chabridon\}@telecom-sudparis.eu}
\thanks{\scriptsize\textcopyright~2026 IEEE. Personal use of this material is permitted. Permission from IEEE must be obtained for all other uses, in any current or future media, including reprinting/republishing this material for advertising or promotional purposes, creating new collective works, for resale or redistribution to servers or lists, or reuse of any copyrighted component of this work in other works. Accepted version; version of record in Proc.\ IEEE CCGrid 2026, pp.~276--286, DOI:~10.1109/CCGrid68966.2026.00036.}
}

\maketitle

\begin{abstract}
Cloud computing is gaining popularity by giving access to seemingly
unlimited virtual resources.  However, Cloud data centres are built
with physical resources and their electricity consumption has been
continuously growing over the past decades.  Microservices are an
important building block of Cloud applications, calling for new
solutions to observe their energy consumption.  Distributed tracing is
widely deployed to diagnose latency and failures in microservice-based
applications, yet it does not expose the energy cost of individual
end-user requests. Such a gap limits energy-aware debugging,
accountability, and control.  This paper presents Spanergy, an
energy-aware distributed tracing approach that correlates
per-microservice power measurements with traces and that attributes
measured energy consumption to request segments, i.e. trace spans. We
showcase Spanergy with synchronous request chains and asynchronous
interactions across microservices.  We present a rigorous experimental
protocol and statistical analysis plan to quantify overhead and to
validate conservation and coverage properties on realistic
configurations. Enabling OpenTelemetry tracing increased total
experiment energy by 59.1\% relative to the uninstrumented baseline,
and Spanergy post-processing added 15.2\% of the baseline
energy. Hence, Spanergy’s incremental energy cost is smaller than the
energy overhead of enabling tracing itself, making the approach
lightweight in practice.  Spanergy also reveals that a non-negligible
fraction of request energy comes from spans outside the
latency-critical path. These results show that energy-aware tracing is
feasible at modest overhead and provides actionable insights for
energy-efficient microservices.
\end{abstract}


\section{Introduction}

Services enabled by Information and Communication Technologies (ICT)
are pervasive and energy-intensive. Estimates place ICT's share of
global greenhouse gas emissions in the \SIrange{1.8}{3.9}{\percent}
range with a likely upward trend~\cite{Freitag2021ICTClimate}. Cloud
data centres alone are projected to increase electricity demand from
\SI{292}{TWh} (2016) to \SI{353}{TWh} by 2030 under conservative
models~\cite{KootWijnhoven2021}, potentially accounting for over
\SI{5}{\percent} of CO\textsubscript{2} emissions by
2030~\cite{Andrae2020}.

Surveys synthesise techniques spanning infrastructure to application
layers for energy efficiency in Cloud
systems~\cite{Mastelic2014,Orgerie2014,Dayarathna2015,KaurChana2015,KhanZakarya2021,Buyya2024}. With
the rise of microservices, recent studies refine taxonomies and
identify opportunities particular to containerised, modular
systems~\cite{Hilman2020,Zhong2022,Araujo2024}. While
microservices can improve resource efficiency, they also introduce
communication overhead and extensive observability stacks that may
affect energy use~\cite{Soldani2018,Haselbock2017}.

In this shift from host and virtual machine controls towards 
software-level observability and attribution, Distributed Tracing (DT) has
matured as a core observability technique to analyse request flows
across services~\cite{Parker2020,Davidson2024}. However, DT tools are
primarily latency-centric, while energy remains an \enquote{odd one
  out} metric with distinct properties and measurement
constraints~\cite{Anand2023OddOne}. DT provides the causal skeleton,
where each request corresponds to one trace that contains one or more
spans linked by context propagation that preserves the exact path,
timing, and concurrency of each request across services. DT does not
yet offer a language- and runtime-agnostic way to map a trace to the
energy it induces: energy is measured at coarse scopes
(host/container), while traces are fine-grained, concurrent, and
distributed across microservices that interact both synchronously and
asynchronously with each other. As a result, practitioners lack (i) a
request-level energy metric that is comparable across endpoints, (ii)
diagnostic guidance on where along a request’s execution energy is
spent, especially when energy-heavy work is not latency-critical. This
gap prevents energy from being treated as a first-class observability
concern, for instance, for operational debugging and trade-off analysis.

In this paper, we propose refining the architecture of traditional
distributed tracing to address the missing link between request
causality and measured joules. Our prototype
implementation ---~Spanergy~--- uses distributed traces as the causal
backbone for end-user requests, and in an offline post-mortem
enrichment stage, it correlates them with per-microservice CPU power
measurements to attribute energy to spans and to entire requests across
both synchronous and asynchronous microservice interactions. The main
contributions of this paper are the following:
\begin{itemize}
\item An energy attribution approach at the request level that enables
  (i)~per-request span attribution, (ii)~diagnostics that can diverge
  from latency-only critical path analysis by revealing
  energy-dominant work outside the latency-critical path, and
  (iii)~endpoint-centric aggregation into a request-level energy
  catalogue that can support decision making, including transfer of energy
  profiles across journeys that reuse the same endpoints.
\item A prototype implementation that shows that distributed tracing
  can be enabled with additional energy cost relative to
  an uninstrumented baseline, and that energy post-processing incurs a
  limited overhead that remains controlled.
\end{itemize}

The remainder of this paper is organised as follows.
First, we present terminology in
Section~\ref{sec:background-distributed-tracing}. Then, in
Section~\ref{sec:motivations}, we motivate the work by formulating
three research questions, and by describing the illustrative microservice-based
application used throughout the paper. 
Sections~\ref{sec:energy-attribution} and~\ref{sec:exp} present the 
approach we propose in Spanergy for energy attribution and then evaluate its
feasibility under realistic microservice interactions, which include
synchronous request chains and asynchronous behaviour. 
Section~\ref{sec:related-work} discusses some work related to distributed
tracing and its consideration of energy consumption.
Section~\ref{sec:conclusion} presents the lessons learned throughout answering
the research questions.

\section{Background on distributed tracing}
\label{sec:background-distributed-tracing}

\begin{figure}[thbp!]
\begin{center}
\includegraphics[width=0.95\columnwidth]{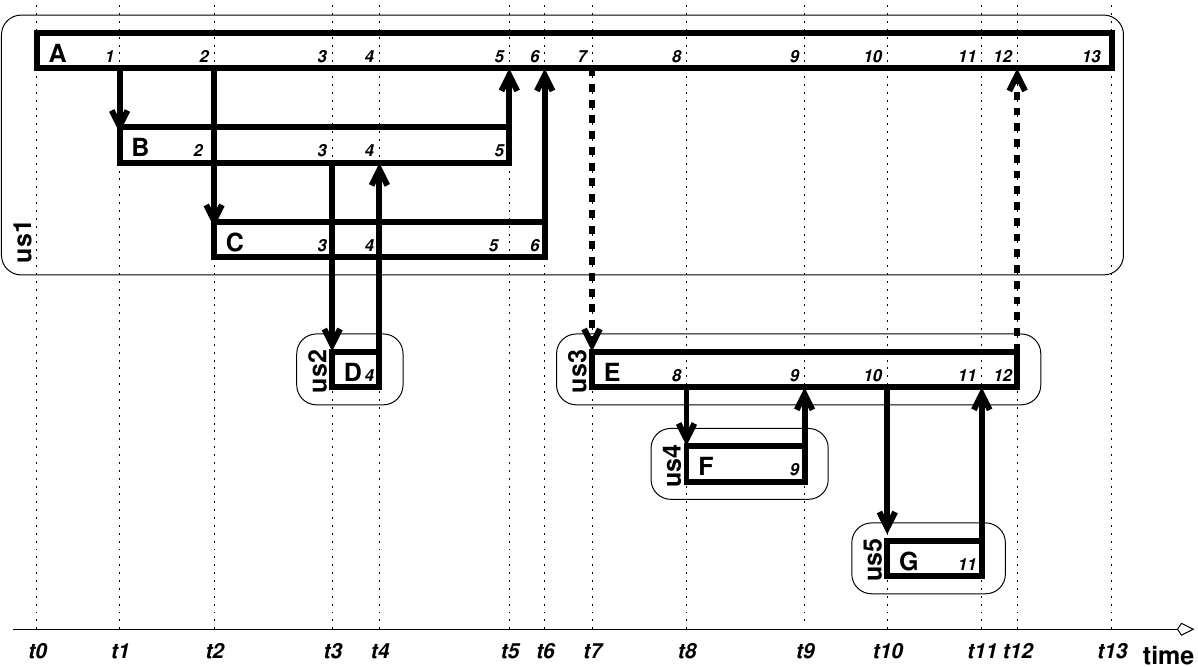}
\caption{Spans of a trace with five microservices ($us_1 \ldots us_5$)}
\label{fig:spans}
\end{center}
\end{figure}

As depicted in Figure~\ref{fig:spans}, the building blocks of
distributed traces are spans
(\code{A}\ldots\code{G})~\cite{otel2024young,otel}. A \textit{span}
represents a unit of work in a microservice. Several spans may
coexist in a microservice, thus showing
concurrent execution (\code{A}\ldots\code{D} in microservice
\code{us1}). For instance, a span is created when a microservice
receives a request and ends before\footnote{For the sake of
completeness, a child span should be created for sending the reply so
that the kind of the request-treatment span is \code{server} and the
kind of the span for sending the reply is \code{client}.} the
microservice sends the replies.  The treatment of the request may
require sending requests to and waiting for replies from other
microservices. The corresponding created spans of the addressees are
linked with the span of the sender by a \code{child} relationship,
which is a causal relationship (line with arrow such as
\code{A}~$\xrightarrow{\text{~~~~~}}$~\code{B}). Clearly, when the
sender is waiting for the replies, it may consume much less energy. In
the case of asynchronous interaction, this corresponds to consuming a message, 
which leads to the creation of a span. Since the producers and
the consumers may not know each other because there exists a
publish-subscribe system between them, the link is called a
\textit{span link} or \textit{follows-from} link in OpenTelemetry 
terminology (dashed line with arrow such as
\code{A}~$\xdashrightarrow{\text{~~~~~}}$~\code{E}). These are the two
causal relationships that we target in this paper.

When a request comes from an end user, we call it an \textit{end-user request}.
The first span of an end-user request is tagged with a unique trace
identifier and has no parent span, i.e. it is a root span, and the span
and trace identifiers are part of the context transmitted to
\code{child} spans. Other kinds of root spans exist in the OpenTelemetry
ecosystem, e.g. the traces that start when consuming a message from a
publish-subscribe system or the traces that start when a batch processor
starts execution. In this work, we ignore batching and restrict our
analysis to spans that (i) belong to the directed acyclic graph (DAG)
induced by the parent--child relationship of the end-user
request, and (ii) belong to sub-traces that are causally connected to that
request through OpenTelemetry \emph{span links} (often rendered as
\code{follows-from} in visualisation tools) with the end-user request's
trace, where the propagated context carries the producer's span identifier.

\section{Motivations and Objectives}
\label{sec:motivations}

In Section~\ref{subsec:research-questions}, we 
formulate the three research questions that motivate this
study. Then, in Section~\ref{subsec:illustrative-application}, we
present the illustrative microservice-based application used
throughout the paper.

\subsection{Research questions}
\label{subsec:research-questions}

In this paper, we motivate and study the following research questions.

\paragraph{Per-request energy estimation}

We target application-level
design and operation decisions with awareness of infrastructure-level 
interactions. The scope of this work is
request-level energy attribution.

Per-request energy estimation is not provided by mainstream tracing
platforms, which model requests as DAGs of spans for latency
diagnosis, nor by existing energy tools that report host, container,
or process-level power without request causality. A principled method
that fuses distributed traces with power meters would enable
attribution of measured joules to individual requests, supporting
per-request budgeting, energy-aware debugging, and comparative
evaluation of designs. In this context, we propose the following
research question: 
\textbf{(RQ1)} How to estimate energy consumption of
individual end-user requests?

\paragraph{Latency \textit{vs.} energy analysis}

The relationship between latency and energy is complex and often
nonlinear. A span with short duration may consume disproportionate
energy if it triggers intensive computation or data
movement. Conversely, a long-running span may consume minimal energy
if it spends most of its time waiting on I/O or network communication.
The latency-critical path pinpoints the longest causally dependent
chain that bounds end-to-end response
time~\cite{MysteryMachine2014,CRISP2022}. A latency-based analysis
would then flag a given span only if it lies on the critical
path. Similarly, latency slack quantifies tolerance to
delay~\cite{MysteryMachine2014}, yet it does not track
energy under overlap. By attributing measured per–service power to
active spans across segments, we obtain span energies that reflect
resource usage during overlap. This enables an energy-critical path
that can diverge from the latency-critical path and expose spans that
dominate joules even when they do not constrain end–to–end latency.

Energy attribution also enables a finer-grained understanding of
resource efficiency. Modern data centres employ heterogeneous hardware
with varying power characteristics. A span executing on a
high-performance processor may complete faster but consume more energy
than the same span on a lower-power core. Latency alone cannot
distinguish between these scenarios. Energy attribution quantifies the
trade-off, allowing operators to make informed decisions about
workload placement. For example, if a span has positive latency slack,
it can be migrated to a more energy-efficient processor without
affecting end-to-end latency. This type of optimisation requires
explicit energy measurement and cannot be inferred from timing data.

The energy-critical path introduced in this work parallels the 
latency-critical path but highlights a different set of spans.  The
energy-critical path surfaces the chain that accumulates the greatest
joule cost, even when its spans are overlapped and thus invisible to
latency reductions. Therefore, latency-critical spans determine
response time, while energy-critical spans dominate energy
consumption. In many cases, these paths diverge. A request may have a
latency-critical path that involves lightweight coordination spans
with minimal energy footprint, while the bulk of energy is consumed by
off-latency-path spans performing data processing or external API
calls.

Therefore, a comparative study of latency vs energy diagnostics
quantifies the added value of energy-aware traces, and we formulate
the following research question: \textbf{(RQ2)} 
To what extent does request-level energy attribution from distributed
traces yield diagnostic insights that differ from latency-only
critical path analysis?

\paragraph{Request-level energy catalogue}

A request-level energy catalogue makes energy attributable to the
application’s externally visible endpoints, enabling stakeholders to
reason about energy in terms of the same API surface used to design,
operate, and govern the system. The catalogue supports concrete
questions that cannot be answered from aggregate host or service
totals, such as which endpoints explain the largest share of total
energy under a given demand mix, and which portions of a journey
concentrate energy. This attribution enables prioritisation that is
directly actionable, because it identifies where energy is spent in
the request space rather than only in infrastructure-level signals.

The catalogue is also useful because it can generalise energy insights
beyond a single reference journey. If endpoints act as reusable
building blocks across multiple journeys, then energy profiles
estimated for those endpoints in a controlled reference journey can be
transferred to other journeys that reuse the same endpoints,
supporting forecasting, budgeting, and governance. This matters for
diverse stakeholders. End users can be informed about the relative
energy cost of actions and can choose lower energy alternatives when
available. Operators and architects can identify energy-dominant
request classes and investigate whether system 
activity or
asynchronous continuations dominate energy. Product owners and
sustainability officers can allocate energy cost to features and track
whether energy reduction efforts target the endpoints that explain
most of the measured energy.

We formulate the following research question: \textbf{(RQ3)}
How useful can a request-level energy catalogue be for supporting
decision making across diverse stakeholders (end-users, operators,
architects, product owners, sustainability officers) with respect to
generalising energy profiles estimated from a reference user journey
to other journeys that reuse the same endpoints?

\subsection{Illustrative application}
\label{subsec:illustrative-application}

\begin{figure}[thbp!]
\begin{center}
\includegraphics[width=0.95\columnwidth]{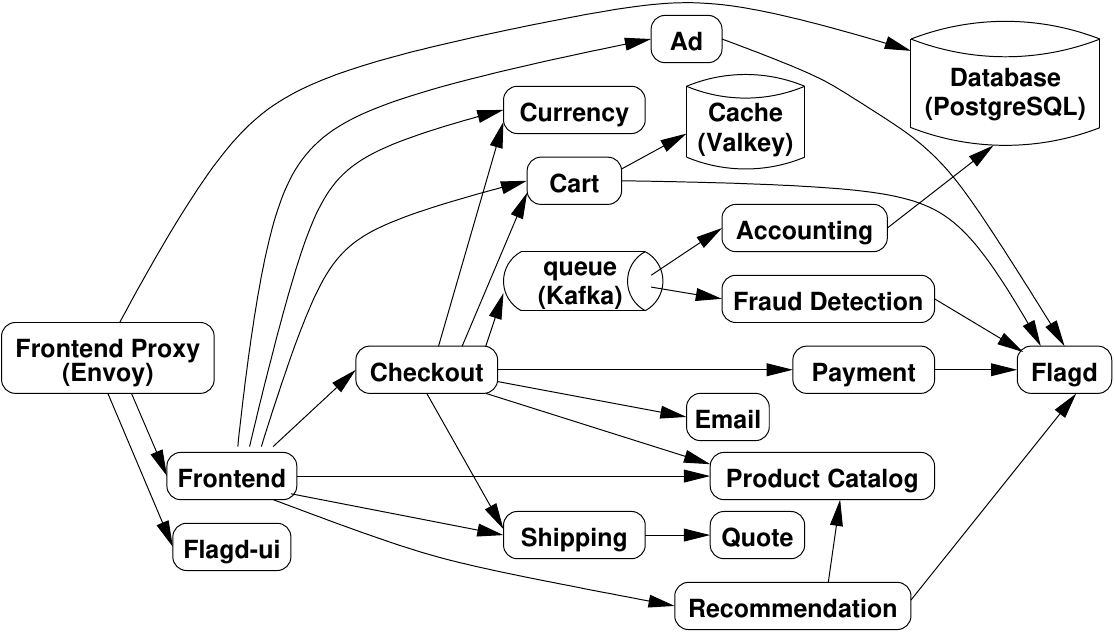}
\caption{Illustrative application: The OpenTelemetry Demo}
\label{fig:illustrative-application}
\end{center}
\end{figure}

Our experiment architecture is based on the microservice
multi-language OpenTelemetry Demo (\texttt{otel-demo})
application\footnote{\url{https://github.com/open-telemetry/opentelemetry-demo/tree/2.1.3}}
that is representative of observability instrumented microservice-based
applications deployed into Clouds. Figure~\ref{fig:illustrative-application}
displays the entities of the application. The application includes a
frontend, microservices built with different frameworks, a
publish-subscribe system (\code{Kafka}), configuration components
(\code{flagd}), and a database (\code{PostgreSQL}). Note that, for
reproducibility reasons, we have excluded the \code{LLM} and the
\code{Product Reviews} microservices because setting up an underlying LLM model
or connecting to an external platform could confound energy measurements.

\section{Energy-aware Distributed Tracing}
\label{sec:energy-attribution}

In Section~\ref{subsec:archi_overview}, we present an overview of the
architecture we have designed for energy-aware distributed tracing.
Then, in Section~\ref{subsec:energy_attr}, we describe the process
followed by Spanergy for attributing energy to
application spans. Finally, in Section~\ref{subsec:lcp-ecp}, we define
the notion of energy-critical path and differentiate it from the
traditional concept of latency-critical path.

\subsection{Architecture for energy-aware distributed tracing}
\label{subsec:archi_overview}

\begin{figure}[t!]
\begin{center}
\includegraphics[width=1.0\columnwidth]{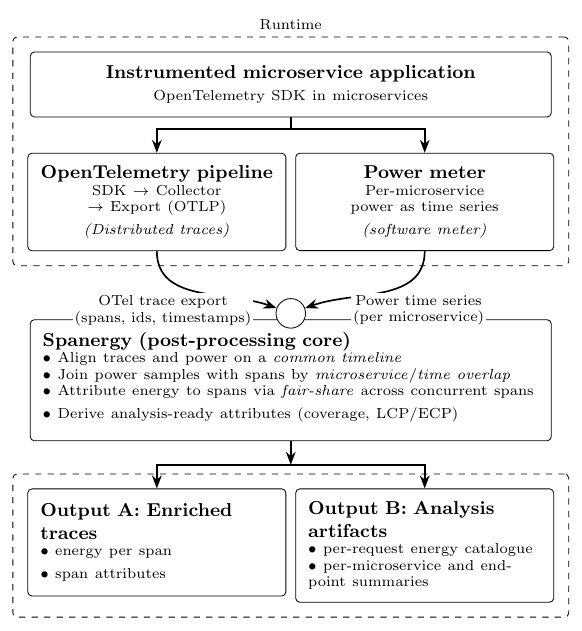}
\caption{Spanergy architecture overview}
\label{fig:spanergy_arch}
\end{center}
\end{figure}

Recent research has been advancing on experimental frameworks for
modelling energy-efficient microservices. Still, there is a gap
towards attributing energy consumption along service call
chains~\cite{Legler2025}. Such a gap is directly addressed by
Spanergy's architecture illustrated in Figure~\ref{fig:spanergy_arch}.
Services are instrumented with OpenTelemetry (OTel) SDKs and export
spans via OTLP to the OTel Collector, which decouples instrumentation
from the choice of trace backend. OpenTelemetry is adopted because it
is vendor-neutral, widely available across languages, and provides
standard processing mechanisms (sampling, batching, compression) to
control telemetry cost in a portable manner.

Spanergy correlates two time-stamped inputs produced during runtime: 
(i) distributed traces, carrying span identifiers and
timestamps, and (ii) per-microservice power time series measured at the
process/container scope. Power and traces are recorded in parallel,
and later aligned. Spanergy operates offline as post-mortem processing: it reads exported
traces and power series, aligns them in time, and assigns measured
service energy to spans.
Although the architecture is agnostic to the choice of software
power meter, the current implementation relies on
Scaphandre\footnote{\url{https://github.com/hubblo-org/scaphandre}} to
collect CPU power at process
scope before mapping those observations to microservice identifiers.
The attribution stage then aligns the resulting power series with
trace timestamps.

\begin{figure}[t!]
\begin{center}
\includegraphics[width=1.0\columnwidth]{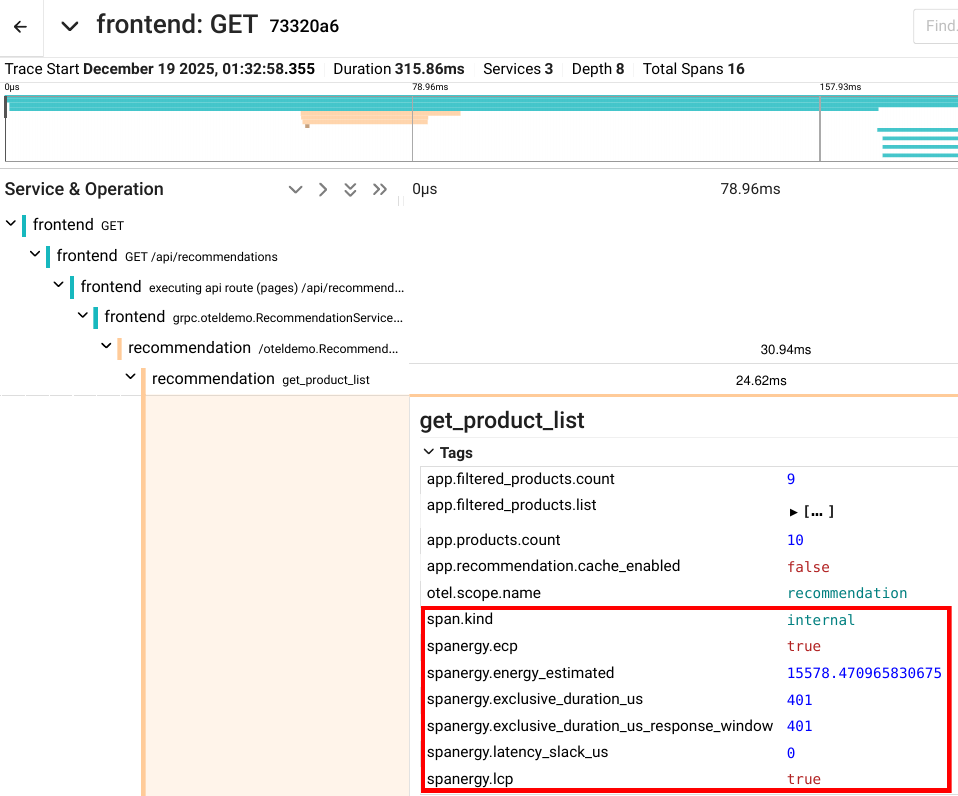}
\caption{Example of an enriched trace visualised using Jaeger}
\label{fig:jaeger_trace}
\end{center}
\end{figure}

Spanergy outputs (A) enriched traces where each span carries an energy
estimate, and (B) analysis artefacts including per-request catalogues,
per-microservice and endpoint summaries, and sanity reports containing
conservation and coverage checks. The enriched traces remain fully
compatible with open-source trace visualisation tools such as Jaeger
(Fig.~\ref{fig:jaeger_trace}), where a distributed trace enriched with
Spanergy attributes can be observed. This compatibility matters
because it enables energy-aware inspection and debugging within
existing observability workflows, without requiring a custom viewer or
a new storage backend.

\subsection{Energy Attribution}
\label{subsec:energy_attr}

In microservice-based applications, several requests can hit the same
microservice at the same time, leading to more concurrent
spans. Commonly, power is measured for the service as a whole, not per
request. The goal is therefore to attribute only the energy actually
produced and to split that energy fairly among concurrent spans. In
our work, we build tooling---Spanergy---to attribute energy estimates
to spans that execute under concurrency: Given per-microservice
stepwise power and a set of spans with start/end times, assign
non-negative energy estimates to spans such that (i)~the sum equals
power integrated over intervals with concurrent spans; (ii)~overlap never
multiplies energy; (iii)~concurrent spans share instantaneous energy
equally; (iv)~it is computed with a single sweep over sorted
boundaries.

To build Spanergy, we have to identify assumptions and restrictions to
the problem. First of all, we assume that all the timestamps are in
the same unit and epoch. During a given time interval (e.g. time
interval $d=[t_4,t_5)$ of Figure~\ref{fig:spans}), we treat power as
constant and equal to the left sample (here, power $p_d$ is equal to
the power at $t_4$). This is conservative and matches how sampled
signals are usually handled. As already noted, time intervals are
right-open to avoid double-counting the boundary. In addition, when
$n$ spans coexist, each receives $1/n$ of the instantaneous
energy.  Spanergy attributes the raw service-scope CPU power reported
by the software power meter in each time slice.  Hence, if no span is
active on a slice, that slice remains unattributed. If spans are
active, the slice energy is shared among them. The current allocator
uses equal-share splitting across concurrent spans to ensure
deterministic service-scope conservation.

Intuitively, as depicted in Figure~\ref{fig:spans}, we cut the
timeline whenever anything relevant changes (beginning or end of a
span) and we only charge energy to the concurrent spans of a
right-open time interval. Since the microservice power is constant
during a right-open time interval, the energy in that time slice is
power times duration, and each span receives a fair share. Summing a
span’s shares over all slices yields its energy, and summing across
the spans of a trace yields the request energy.

Practically, we normalise microservice identifiers so that traces and
power samples match without ambiguity. Instead of checking each span
against every energy
interval, we create one sorted list of all
relevant boundaries (span starts/ends and power changes) and traverse
it once. We keep the set of concurrent spans up to date and attribute
energy slice by slice. In addition, for the computation of the set of
concurrent spans at a given instant $t$, we encode the right-open
semantics by removing, from the active set, the spans that end at $t$, and then
adding the spans that start at $t$. Finally, we cut the last energy
interval at the last span end. This avoids long idle tails that would
complicate accounting and audits.

Therefore, the energy attribution algorithm can be sketched as
follows:
\begin{enumerate}
  \item Collect boundaries: Take all span starts, span ends,
    and power-change times; sort and de-duplicate.
  \item Apply the tie-break at each boundary: First remove spans that
    end, then update power if it changes, then add spans that start.
  \item Form the next slice: The current boundary and the
    next boundary define a half-open slice of time.
  \item Attribute the slice: If at least one span exists,
    distribute the slice energy ($\text{power} \times \text{duration}$)
    equally among the active spans.
  \item Repeat to the end: Continue until the last span end for the
    service; remove that last span to avoid long idle tails that would
    complicate accounting; write the assigned energy back to spans.
\end{enumerate}

\subsection{Latency and energy critical paths}
\label{subsec:lcp-ecp}

As depicted in Figure~\ref{fig:spans}, a span is decomposed into span
fragments. The \emph{latency-critical} path is defined as the chain of span
fragments that determine the overall response time: If any span
fragment on this chain completes earlier, the entire request completes
earlier by the same amount~\cite{MysteryMachine2014,CRISP2022}. Each
span fragment has a duration and we can compute the part of this
duration that has no concurrent span fragments from child spans. The
latency-critical path is found by starting at the first fragment of
the root span and repeatedly selecting the child span fragment that
finishes last after accounting for non-overlapping children.
For instance, in Figure~\ref{fig:spans}, the latency-critical path is
computed as the following segments: $A_{1\text{--}2} + C_{3\text{--}6}
+ A_7 + E_8 + F_9 + E_{10} + G_{11} + E_{12} + A_{13}$.

\begin{figure}[t!]
\begin{center}
\includegraphics[width=0.95\columnwidth]{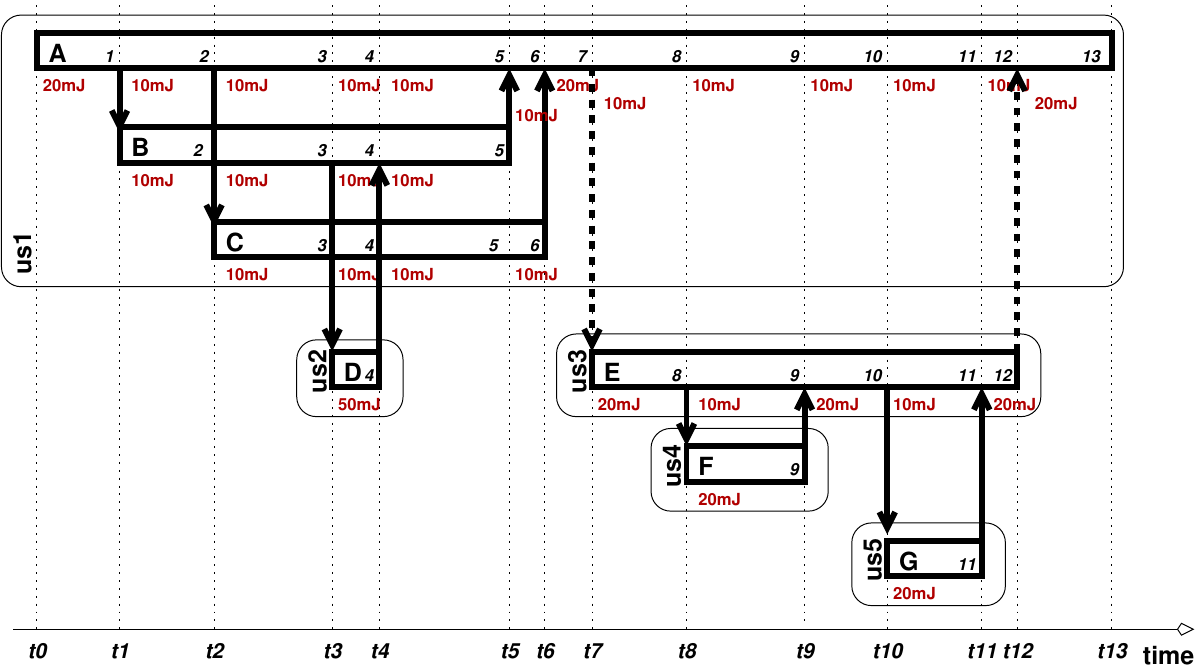}
\caption{Span fragments with energy consumption attribution}
\label{fig:spans_with_energy}
\end{center}
\end{figure}

Once energy has been attributed to spans, we can study the
distribution of energy across the trace. By considering the graph of a
trace built from \textit{child} and \textit{follows-from} relationships between
spans and with spans decomposed into span fragments, the
\emph{energy-critical} path is the root-to-leaf path maximising the sum of
per-span energies. In Figure~\ref{fig:spans_with_energy}, we
complement Figure~\ref{fig:spans} with energy consumption
attribution. The energy-critical path, that is the path that consumes
the most energy, is $A_1 + B_{2\text{--}3} + D_4 + B_5 + A_6 + E_8 +
F_9 + E_{10} + G_{11} + E_{12} + A_{13}$ with $250mJ$.

\section{Experimental evaluation}
\label{sec:exp}

This section is organised as follows. First, we specify, in
Section~\ref{subsec:testbed}, the experimental design used to validate
request-level energy attribution. We present the scenarios and
experimentation protocol in Section~\ref{subsec:scenarios}. Then, in
Section~\ref{subsec:metrics}, we detail the metrics used to quantify
the incremental energy overhead of Spanergy when moving from scenario
S0 with no tracing, to S1 with tracing only, to S2 with tracing plus
Spanergy. Next, we describe the validation process in
Section~\ref{subsec:validation} and the hypotheses for statistical
analysis in Section~\ref{subsec:hypotheses}. Finally, in
Sections~\ref{subsec:results} and~\ref{subsec:threats}, we present the
results and discuss threats to validity.

The system under test is the OpenTelemetry Demo application, referred
to as \texttt{otel-demo}. This application has been selected because
it includes synchronous request chains and asynchronous interactions
across microservices.

The design follows guidance for controlled experimentation and
statistical conclusion validity in empirical software
engineering~\cite{Wohlin2012Experimentation,ArcuriBriand2014Hitchhikers,KitchenhamMadeyski2017Robust}. The
acceptability question is framed as an equivalence style comparison,
which is appropriate when the objective is to show that an overhead
remains below a practical threshold~\cite{Dolado2014Equivalence}.
The analysis can be further assessed using the replication
package at \url{https://zenodo.org/records/18677258}.

\subsection{Testbed and deployment}
\label{subsec:testbed}

Experiments are executed on the Grid'5000\footnote{\url{https://www.grid5000.fr}} site in Lyon using the
Taurus cluster. Each run reserves one node and starts from a
clean operating system image provided by the testbed deployment
mechanism. The Taurus nodes in this study have the following
configuration. Model is Dell PowerEdge R720. CPU is Intel Xeon E5-2630
Sandy Bridge at 2.30GHz with 2 CPUs per node and 6 cores per
CPU. Memory is 32 GiB.

The \texttt{otel-demo} application is deployed using Docker
Compose. Locust\footnote{\url{https://locust.io/}} is used as the external load generator. The workload
is fixed across all the runs triggered by the same Locust script and the
same deterministic request sequence, with a fixed total of 6{,}410
requests that generate internal synchronous and asynchronous
microservice interactions. The request mix represents a user journey
that includes browsing, cart operations, and checkout, plus auxiliary
endpoints. The exact request mix and sequencing are specified by the
load generator artefact released with the reproduction package. 

\subsection{Scenarios and experimentation protocol}
\label{subsec:scenarios}

We define three scenarios:
\begin{itemize}
\item \textbf{S0, no tracing.} The application executes with tracing
disabled at the SDK level and no spans are exported.
\item \textbf{S1, tracing only.} The application executes with OpenTelemetry
tracing enabled using a fixed configuration, with sampler and exporter
settings held constant across runs. This scenario captures the energy
overhead attributable to tracing, exporting and the presence of the OpenTelemetry
Collector.
\item \textbf{S2, tracing plus Spanergy.} The load phase is identical to S1
and produces the same observability data. After the load completes,
Spanergy post-processing executes to compute request-level energy
attribution outputs. The energy consumed by this post-processing is
treated as Spanergy overhead beyond tracing.
\end{itemize}

The protocol produces 60 runs, and each run includes warm-up and
cool-down intervals. First, 30 S0 runs are executed
sequentially. Then, 30 combined runs are executed sequentially. Each
combined run contains Phase~1, which is the load phase under tracing,
and Phase~2, which is the Spanergy post-processing phase executed
after Phase~1. In these combined runs, the S1 metric corresponds to
the total energy from Phase~1. The S2 metric corresponds to the total
energy from Phase~1 plus the energy consumed during Phase~2.

\subsection{Metrics, instrumentation, and energy computation}
\label{subsec:metrics}

The primary metric for the overhead study is the total experiment
energy. For a run, the total experiment energy is computed as the sum of
energies across all the \texttt{otel-demo} microservices included in the
experimental scope. Process-level power samples are collected and
mapped to microservices. For each microservice, samples are treated as a time
ordered power series with power in watts and timestamps in seconds.
Power telemetry is collected with Scaphandre, which estimates per-process CPU power from
Linux powercap/RAPL energy counters and \code{/proc} CPU-time
accounting between consecutive scrapes. During export, container- and
service-level power are obtained by aggregating the corresponding PIDs
into service-scope series used by the attribution workflow.

Energy and overhead
metrics are computed on the measurement window that excludes
deployment transients.
Energy for a microservice is computed using trapezoidal integration over
consecutive power samples. For a microservice $us$ with power samples $(t_k,
P_k)$ ordered by time, energy is computed as
\begin{equation}
E_{\mathrm{us}} = \sum_{k=2}^{n} \frac{P_{k-1} + P_k}{2} \times (t_k - t_{k-1})
\label{eq:trapz}
\end{equation}
where $(t_k - t_{k-1})$ is the elapsed time in seconds and
$E_{\mathrm{us}}$ is in joules. If a microservice has only one power
sample in the measurement window, trapezoidal integration is not
defined. In this rare data quality case, energy is approximated by
assuming constant power over a fallback duration, which is computed as the median of
per-target median sampling intervals over targets with at least two samples.
In this campaign, the median per-microservice power sampling
interval is about 2.12s. Span durations are mostly sub-second, with a
median around 27ms and P90 around 171ms. This mismatch can bias very
short spans. For this reason, the analysis emphasises run-level and
endpoint-level aggregates with confidence intervals and interprets
fine-grained span values together with coverage diagnostics.

Total energy for a run is computed as
\begin{equation}
E_{\mathrm{tot}} = \sum_{\mathrm{us} \in \mathcal{S}} E_{\mathrm{us}}
\label{eq:Etot}
\end{equation}
where $\mathcal{S}$ is the fixed set of microservices considered for
aggregation. 

In addition to energy, the study records latency and throughput from
Locust, CPU and memory utilisation from host and process metrics, and trace
volume (number of exported spans during Phase~1 within the measurement window)
from exported telemetry artefacts.

\subsection{Validation}
\label{subsec:validation}

Request-level energy attribution is validated
against energy conservation at the microservice horizon.
For each microservice $us \in \mathcal{S}$, let $E_{us}$ be the
measured microservice energy from Eq.~\ref{eq:trapz} over the load window,
and let $E^{\mathrm{attr}}_{us}$ be the sum of Spanergy attributed
energies over all the spans executed by $us$ within the same window. The
analysis reports the conservation ratio $r_{us} = E^{\mathrm{attr}}_{us} /
E_{us}$. Ratios above~1 indicate over attribution and must not occur
beyond a small tolerance that accounts for numerical and alignment
error. The analysis flags microservices with $r_{us} > 1.01$ and reports the
maximum observed violation per run. The analysis also reports $1-r_{us}$
to quantify energy that is not covered by spans, which captures
background activity and gaps in tracing coverage.

\subsection{Hypotheses and statistical analysis}
\label{subsec:hypotheses}

Let $E_{\text{S0},i}$ denote total energy for S0 run $i$ with $i \in
\{1,\dots,30\}$.
Let $E_{\text{S1},j}$ denote S1 energy corresponding to Phase~1 energy in combined run
$j$ and let $E_{\text{Sp2},j}$ denote Phase~2 energy in the same run, with $j
\in \{1,\dots,30\}$. The S2 energy for combined run $j$ is then $E_{\text{S2},j} =
E_{\text{S1},j} + E_{\text{Sp2},j}$.

Let $\mu(\cdot)$ denote the mean of the underlying distribution of per-run
energy for a scenario under the fixed workload.
Sample means are denoted by $\bar{E}$.
This sample size supports stable $95\%$ confidence
interval estimation for overhead metrics.

\paragraph{Tracing overhead, S0 to S1}
The first hypothesis quantifies the mean energy increase when tracing
is enabled.
\begin{align}
H_{0,01} &: \mu(E_{\text{S1}}) \le \mu(E_{\text{S0}}) \\
H_{1,01} &: \mu(E_{\text{S1}}) > \mu(E_{\text{S0}})
\end{align}
Since S0 runs and S1 observations come from different runs, inference
uses independent sample procedures. The analysis reports the
difference in means $\mu(E_{\text{S1}}) - \mu(E_{\text{S0}})$ and the relative
overhead $(\mu(E_{\text{S1}}) - \mu(E_{\text{S0}}))/\mu(E_{\text{S0}})$, each with $95\%$
confidence. Normality is assessed using the Shapiro-
Wilk test~\cite{ShapiroWilk}. Welch procedures~\cite{Welch} are used as the primary parametric
reference due to unequal variance robustness.

Practical significance is reported using effect
sizes~\cite{ArcuriBriand2014Hitchhikers}. The analysis reports Hedges
$g$ for the standardised mean difference with small sample correction
and Cliff's delta for ordinal dominance. Hedges $g$ is interpreted
using standard conventions for small, medium, and large
effects. Cliff's delta is interpreted using the standard magnitude
thresholds reported in the literature.

\paragraph{Spanergy acceptability relative to tracing}
The second hypothesis evaluates whether Spanergy overhead remains
below the tracing overhead, using $\mu(E_{\text{S0}})$ as the common
baseline. The criterion is $\mu(E_{\text{Sp2}}) \le \mu(E_{\text{S1}}) -
\mu(E_{\text{S0}})$. This yields
\begin{align}
H_{0,A} &: \mu(E_{\text{Sp2}}) \ge \mu(E_{\text{S1}}) - \mu(E_{\text{S0}}) \\
H_{1,A} &: \mu(E_{\text{Sp2}}) < \mu(E_{\text{S1}}) - \mu(E_{\text{S0}})
\end{align}
The analysis reports the relative Spanergy overhead
$\mu(E_{\text{Sp2}})/\mu(E_{\text{S0}})$ with a $95\%$ confidence interval. The
analysis also reports the acceptability statistic
\begin{equation}
S = \mu(E_{\text{Sp2}}) - \left(\mu(E_{\text{S1}}) - \mu(E_{\text{S0}})\right)
\label{eq:accept}
\end{equation}
and a one-sided $95\%$ confidence interval computed using delta
method standard errors that account for the covariance between
$E_{\text{S1},j}$ and $E_{\text{Sp2},j}$ within combined runs. This hypothesis
structure follows equivalence and non-inferiority reasoning used in
experimental software engineering when the objective is to support a
bounded overhead claim~\cite{Dolado2014Equivalence}.

\subsection{Experimental results}
\label{subsec:results}

\paragraph{Power quality assurance}

We seek to establish temporal coherence between Spanergy’s
request-level energy and independent observations.  Concretely, we
assess whether the time profiles \(E_{\text{wall}}(t)\),
\(E_{\text{host}}(t)\), \(E_{\text{app}}(t)\), and
\(E_{\text{trace}}(t)\) co-vary, exhibiting similar rises, falls, and
event timing, even if their absolute magnitudes differ due to meter
scope and granularity. If \(E_{\text{trace}}(t)\) follows the same
time pattern as \(E_{\text{wall}}(t)\), \(E_{\text{host}}(t)\), and
\(E_{\text{app}}(t)\) (high correlation, small lag), this supports
that Spanergy preserves real energy dynamics at the request
level.

\begin{figure}[t!]
\begin{center}
\includegraphics[width=0.95\columnwidth]{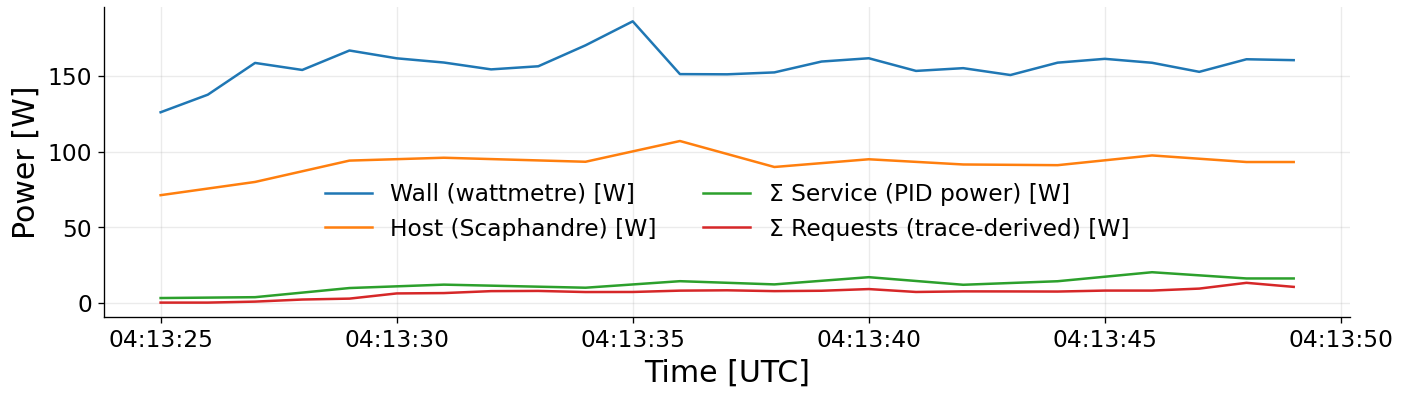}
\caption{Time-aligned power trajectories across measurement levels}
\label{fig:power-over-time}
\end{center}
\end{figure}

\begin{figure}[t!]
\begin{center}
\includegraphics[width=0.95\columnwidth]{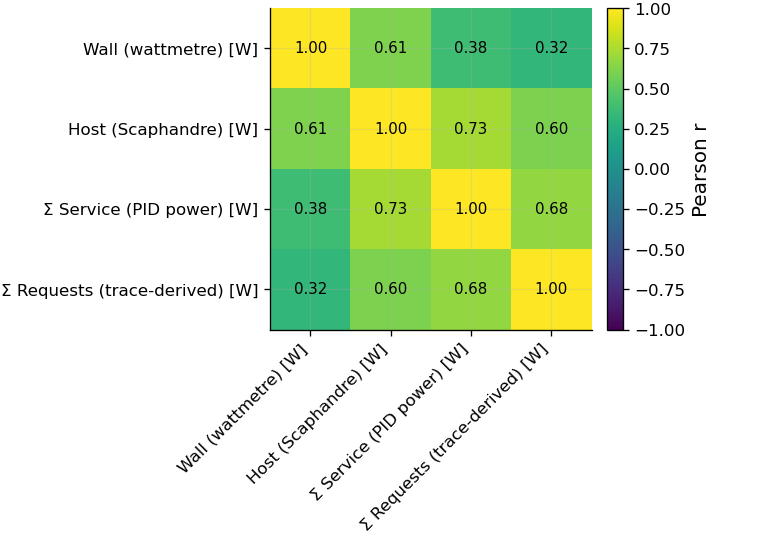}
\caption{Power (Pearson) correlation matrix}
\label{fig:power-matrix}
\end{center}
\end{figure}

For this particular question, we select as representative the run whose Phase~1 
energy is closest to the median across all combined runs.
Figure~\ref{fig:power-over-time} relates power levels through time,
while Figure~\ref{fig:power-matrix} shows the power correlation matrix
for that representative run. The host and
summed service power signals co-vary strongly ($r=0.73$), and the
trace-derived request power tracks both the service aggregate
($r=0.68$) and the host signal ($r=0.60$). Correlations with the wall
meter are weaker ($r\in[0.32,0.61]$), which is consistent with scope
differences and a non-negligible baseline outside the container
set. This indicates that trace-derived
request energy captures a substantial fraction of the measured service
energy dynamics while leaving a remainder attributable to background
activity and incomplete coverage.

\paragraph{Spanergy overhead}

\begin{figure}[thbp!]
\begin{center}
\includegraphics[width=0.95\columnwidth]{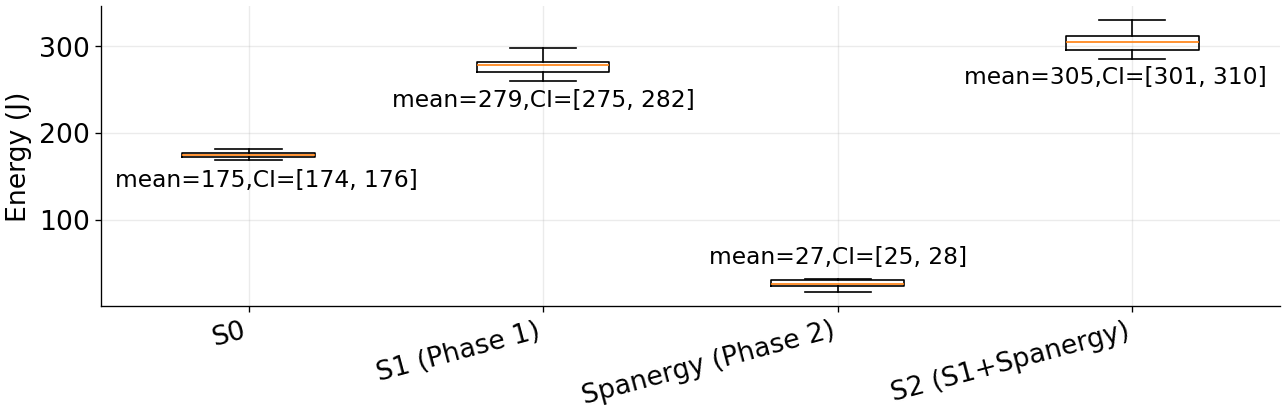}
\caption{Energy consumption across scenarios}
\label{fig:s0-s1-s2}
\end{center}
\end{figure}

The S0--S1 contrast isolates the incremental energy attributable to
tracing, exporting, and the Collector under a fixed workload. Figure
\ref{fig:s0-s1-s2} illustrates energy consumption for the
scenarios. Welch inference on run-level totals indicates a mean
increase of 103.5~J, with a 95\% confidence interval of [99.4,
  107.6]~J and a relative overhead of 0.591 (95\% CI [0.567,
  0.616]). The p-value ($3.1\times 10^{-35}$) provides strong evidence
against $H_{0,01}$. Effect sizes support practical relevance: Hedges
$g=13.1$ indicates a difference far larger than within-scenario
dispersion, and Cliff’s $\delta=1.0$ indicates complete dominance of
S1 over S0 across the observed runs.

The acceptability hypothesis evaluates whether Spanergy
post-processing remains below the tracing overhead budget defined by
$\mu(E_{\text{S1}})-\mu(E_{\text{S0}})$. The mean relative Spanergy overhead is
0.152 of the S0 baseline (95\% CI [0.143, 0.161]). The acceptability
statistic in Eq.~\ref{eq:accept} is negative ($S=-77.0$~J) and its
one-sided 95\% upper confidence bound remains below zero
($S_{0.95}=-73.5$~J), supporting $H_{1,A}$ under the adopted
confidence level. This indicates that the incremental cost of Spanergy
post-processing is substantially smaller than the energy already
incurred by enabling tracing.

\paragraph{Per-request energy estimation}

The following discussion refers to Research Questions~1 and~3.
For RQ3,
outputs are considered useful when they
support reproducible decisions through stable endpoint ranking,
interpretable demand-to-service decomposition, or diagnostic signals.
When computed repeatedly over comparable intervals, those indicators
also form endpoint-level time series that can reveal recurring and
seasonal energy demand patterns.

\begin{figure}[t!]
\begin{center}
\includegraphics[width=0.95\columnwidth]{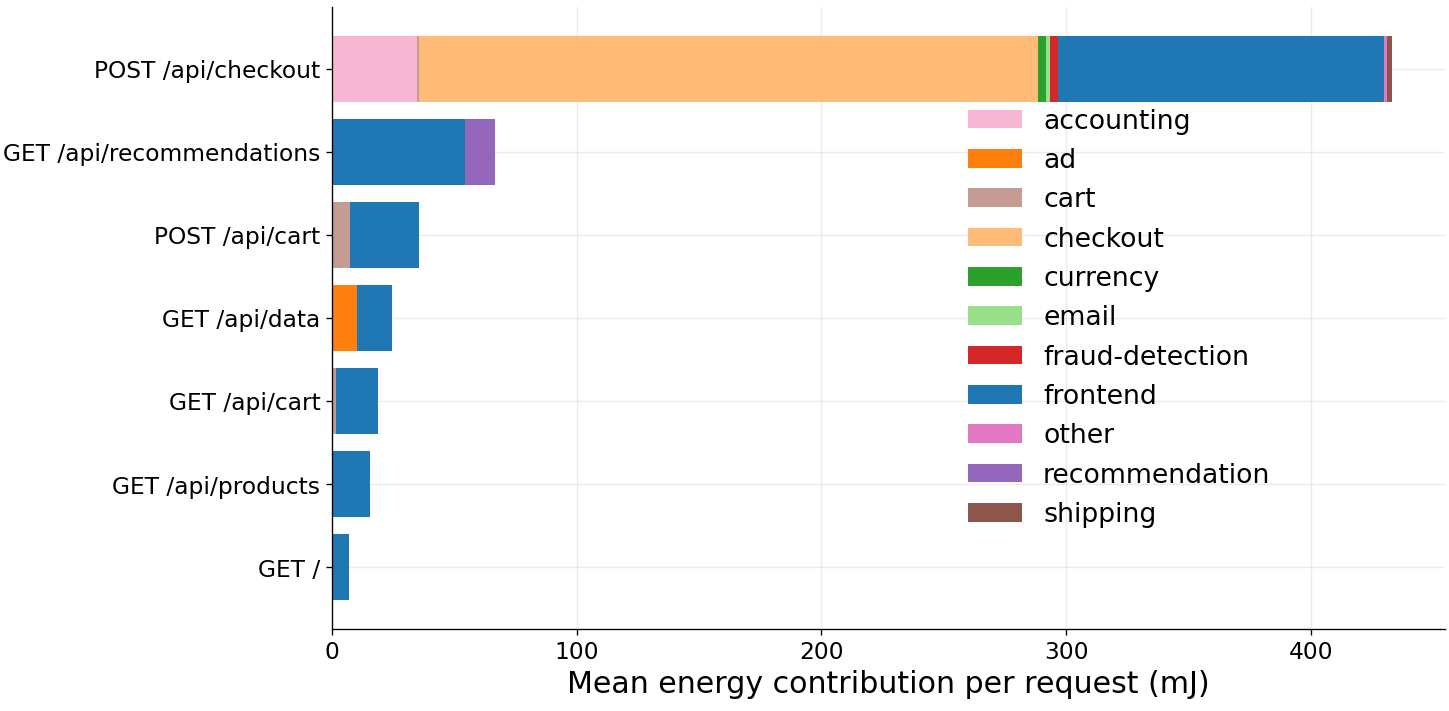}
\caption{Endpoint energy composition by microservice}
\label{fig:energy-endpoint-service}
\end{center}
\end{figure}

Using Spanergy-attributed energies on end-user requests, the
analysis constructs a request energy catalogue that maps canonical
endpoint signatures to mean end-to-end energy and uncertainty across
runs. The resulting footprint is concentrated: \code{POST
  /api/checkout} exhibits a mean of 0.390~J per request (95\% CI
[0.360, 0.417]), while \code{GET /api/recommendations} averages
0.0657~J (95\% CI [0.0642, 0.0672]) and \code{POST /api/cart}
averages 0.0355~J (95\% CI [0.0349, 0.0361]). These endpoint-level
estimates support reasoning at the API boundary and enable “bill of
materials” decompositions of per-microservice estimations into
contributions from dominant request types as
Figure~\ref{fig:energy-endpoint-service} illustrates.

These results show that attribution enables reliable estimation of
end-to-end energy for individual end-user requests, with per-endpoint
means and confidence intervals that are consistent across runs, hence
answering RQ1. In particular, the request-level energy catalogue
exposes clear differences between endpoints such as \code{POST
  /api/checkout} and \code{GET /api/recommendations} that would remain
invisible at host- or service-level totals.
In a SaaS context, these outputs can support role-specific
decisions. Operators can prioritise optimisation on high-joule
endpoints. Architects can compare interaction alternatives under joule
budgets. Product teams can rank journeys by operational energy
footprint. Sustainability teams can monitor endpoint-level progress
against reduction targets. End users can be informed about lower energy
or quality of service alternatives when equivalent actions exist.

\paragraph{Span energy in LCP and ECP}

The following discussion refers to Research Question~2.

Classic latency-critical path (LCP) analysis characterises performance
using the span chain that dominates end-to-end latency. Spanergy
enables a complementary energy-critical path (ECP) view that
focuses on attributed energy within the same request. For each end-user request,
we compute one latency-critical path (LCP) and
one energy-critical path (ECP) on the same trace graph, and treat each
path as the set of spans it selects. We then compute a per-request
Jaccard overlap $|LCP \cap ECP| / |LCP \cup ECP|$, aggregate these values
within each run, and finally report the mean and confidence intervals
across the 30 runs. Across runs, the span-set overlap between LCP and ECP is high
but not identical: the mean per-request Jaccard similarity overlap is 0.872 with
a 95\% CI of [0.869, 0.875], indicating a systematic yet bounded divergence
between latency-dominant and energy-dominant work.

\begin{figure}[t!]
\begin{center}
\includegraphics[width=0.95\columnwidth]{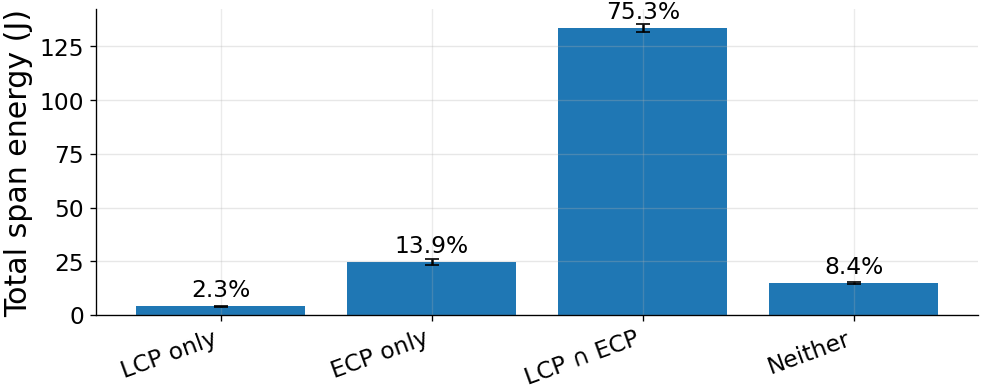}
\caption{Span energy distribution by LCP/ECP membership}
\label{fig:energy-lcp-ecp}
\end{center}
\end{figure}

From an energy perspective, most span-attributed energy lies on spans
that are both latency- and
energy-critical. Figure~\ref{fig:energy-lcp-ecp}
shows that spans in $\mathrm{LCP}\cap\mathrm{ECP}$
account for 75.3\% of span energy (95\% CI [74.8, 75.9]) while
representing 57.7\% of spans by count. At the same time, 13.9\% of
span energy lies on ECP-only spans (95\% CI [13.4, 14.5]), which are
not latency-critical and would be invisible to latency-only diagnoses.
The mean share of request energy on the LCP is 0.861, leaving approximately 14\% of request energy
outside the latency-critical chain. Note that Jaccard overlap is a span-count
similarity metric, whereas Figure~\ref{fig:energy-lcp-ecp}
reports energy shares by class, so the two quantities are informative but not
directly comparable. They differ because Jaccard gives equal weight to spans and
uses only the $LCP \cup ECP$ set, while
Figure~\ref{fig:energy-lcp-ecp} weights by attributed energy and also
includes the Neither category.

These findings answer RQ2 by showing that energy-aware diagnostics
can diverge from latency-only critical path analysis: while most
energy lies on spans that are both latency- and energy-critical, about $14\%$ 
of request energy is concentrated on ECP-only spans that latency-centric methods
would miss. This divergence demonstrates that request-level energy attribution
yields additional diagnostic insights beyond what latency-based critical path
analysis can provide.

\begin{figure}[t!]
\begin{center}
\includegraphics[width=0.95\columnwidth]{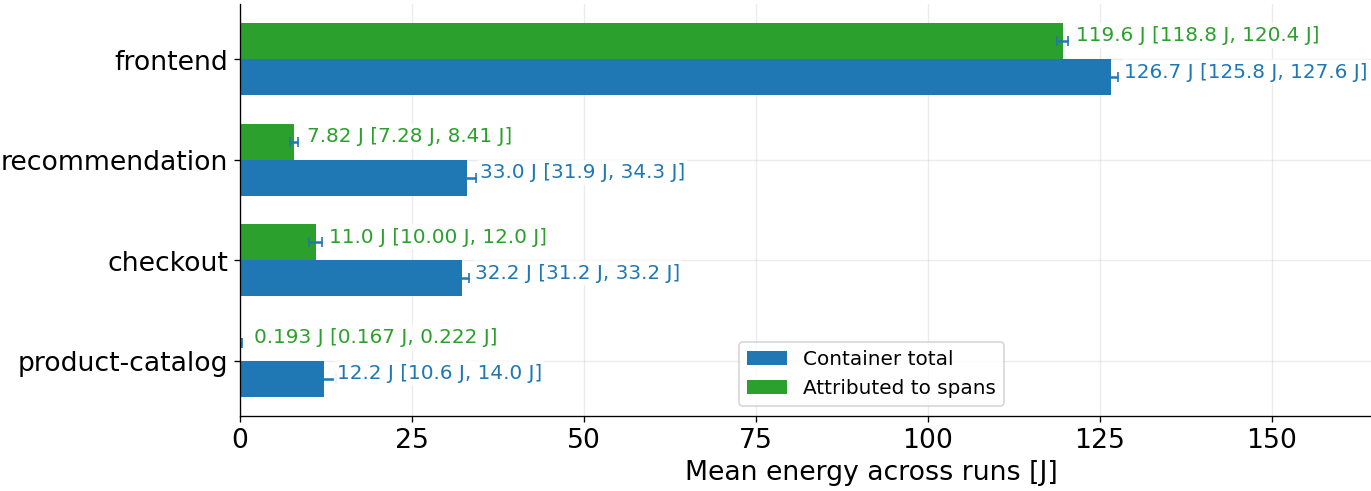}
\caption{Top 4 services — Span-attributed vs container total energy}
\label{fig:top4-campaign}
\end{center}
\end{figure}

\begin{figure}[t!]
\begin{center}
\includegraphics[width=0.95\columnwidth]{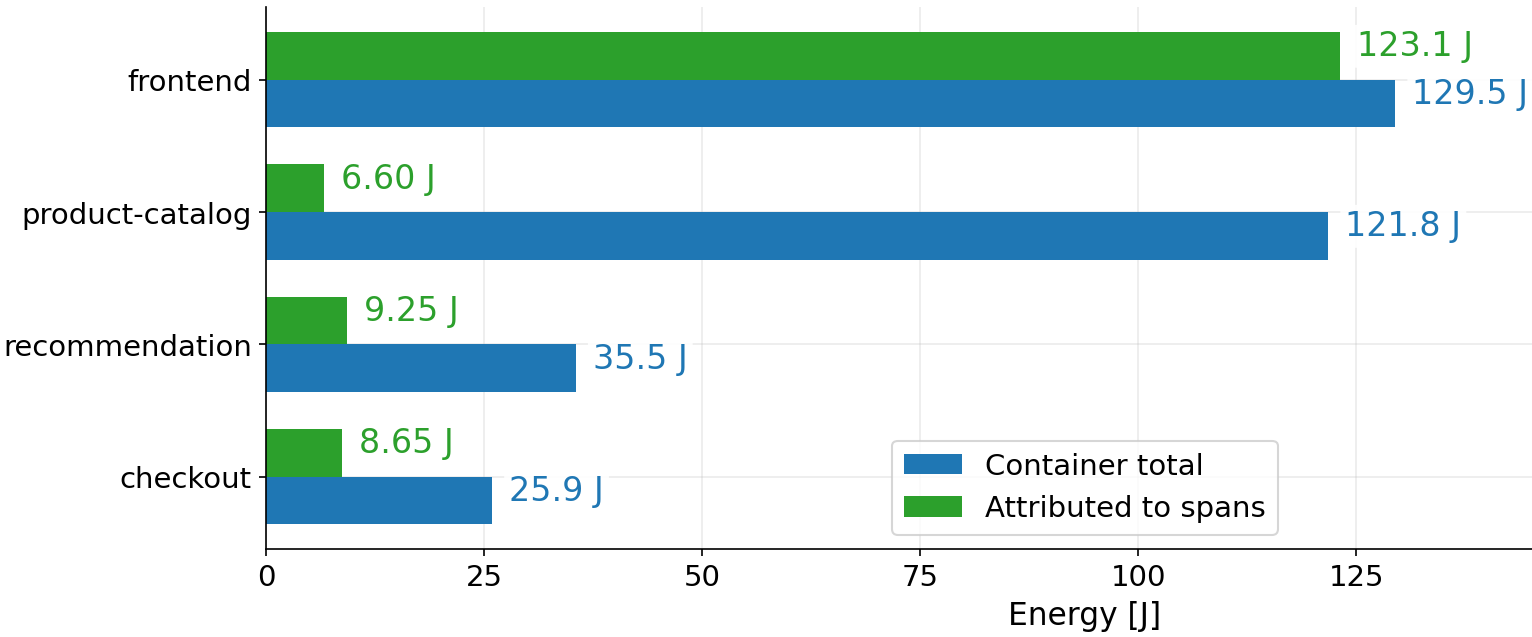}
\caption{Single dark energy run top 4 services — Span-attributed vs container total energy}
\label{fig:top4-dark-energy}
\end{center}
\end{figure}

\paragraph{Dark energy debugging}
We distinguish traced-span attributed energy from the
unattributed remainder at microservice scope. The remainder denotes
measured service energy that is not mapped to traced spans in the
selected window. It can encompass background runtime activity, I/O operations,
incomplete instrumentation coverage, or asynchronous work with missing
or delayed causal links.
During Spanergy’s development and evaluation, we observed that it can
also serve as a practical diagnostic instrument for identifying
anomalous energy behaviour and guiding corrective deployment changes.
Figure~\ref{fig:top4-campaign} contrasts, for the four most
energy-consuming services, the total container energy with the subset
of energy that Spanergy attributes to traced spans. Campaign-wide (30
runs), the \code{product-catalog} container consumes on average 12.2~J,
whereas only 0.193~J are attributed to its spans. In earlier
experiments conducted with the official repository deployment
configuration, prior to stabilising the application’s runtime
behaviour, the \code{product-catalog} container consumed approximately
121.8~J (Figure~\ref{fig:top4-dark-energy}), i.e., an order of magnitude
above the campaign baseline. Interestingly, only a small fraction of this
elevated consumption was explained by span-attributed energy. We refer
to this mismatch as the \emph{dark energy problem}: container-level
energy is high, yet the traced-span energy remains low, leaving a large
unattributed remainder. This observation was particularly unexpected
under our fixed request mix, where \code{product-catalog} is not
expected to experience sustained stress comparable to the
\code{front-end} service, which acts as the application entry point.

Subsequent investigation revealed that \code{product-catalog}
exhibited constant CPU saturation (close to 100\% utilisation), which
plausibly explains the excessive energy consumption. The issue was
mitigated by relaxing memory constraints: increasing the container
memory limit from 20~MB to 80~MB, and the Go runtime memory limit from
16~MB to 64~MB. This configuration change reduced garbage-collection
pressure, alleviated CPU-bound behaviour, and restored the service to
a stable operating regime with normalised performance and energy
consumption.

\subsection{Threats to validity}
\label{subsec:threats}

Internal validity is primarily threatened by temporal misalignment
between power sampling, span timing, and request boundaries. The
analysis mitigates this risk by integrating microservice-level power series
over a consistent measurement window that excludes deployment
transients (Eq.~\ref{eq:trapz}). Residual alignment error, sampling
jitter, and rare fallback cases with a single sample can still bias
short-lived spans and small microservices. Additional internal threats
arise from microservice-to-process mapping and from non-deterministic
background activity within containers, which can inflate energy
occurring outside traced spans and affect catalogue coverage and
conservation checks.
Spanergy mitigates interpretation risk
through explicit conservation checks with conservation ratio $r_{us}$, uncovered-energy
reporting with $1-r_{us}$, and coherence checks against independent
power signals. These controls expose uncertainty and support auditing.
However, they do not eliminate attribution imprecision.

Construct validity concerns whether our measurements and
operationalisations faithfully capture the intended constructs (here,
request/span energy attribution and LCP/ECP membership). Construct validity
depends on trace completeness and on the correctness of request and span
classification. Sampling strategies
that drop spans reduce attribution coverage, and tail-based strategies may
introduce bias into traces that are retained. Although crucial in production
systems, this study does not apply any sampling, therefore reports
coverage diagnostics and treats catalogue-based inference as
conditional on sufficient coverage.
Construct validity is also affected by the equal-share overlap
rule used in this study. When concurrent spans have heterogeneous
resource intensity, per-span allocation can shift even when
service-level conservation holds.
This motivates resource-weighted overlap allocation as an important
future direction.
This type of validity can be further affected by the power basis used
for attribution. The current implementation allocates raw
service-scope CPU power reported by the software power meter, without
subtracting a per-service idle baseline.
Consequently,
span energies in active windows reflect
both dynamic and baseline components, so dynamic-only interpretations
should be framed accordingly, especially under low utilisation or changing
concurrency.

External validity is limited by the chosen workload, deployment
configuration, and hardware. The campaign uses a single deterministic
Locust script, a fixed tracing configuration, and one cluster
type. Overhead magnitudes and catalogue values may change under
different request mixes, scaling regimes, or tracing
configurations. Statistical conclusion validity further relies on
run-level independence and on the stability of the experimental
environment across sequential run blocks. Homogeneous nodes and clean
deployments reduce confounding, but unobserved drift across time
remains a plausible source of residual variance.

Scalability is a key quality attribute in real-world microservice
systems that may comprise thousands of services and complex interactions.
Because Spanergy attributes energy only to retained spans, scalability is
directly coupled to sampling policy. At large scale, sampling should be
treated as a constrained estimation design that preserves endpoint
representativeness while bounding telemetry ingestion and storage cost.
A practical direction could be to combine adaptive tail sampling with endpoint
stratification so that low-frequency high-energy traces remain observable
alongside common traffic. Under this regime, each reporting window should
publish coverage and conservation confidence indicators to assess result
stability. In this context, the current full-tracing campaign provides a
reference point for attribution overhead and coverage before sampling is
introduced. The proposed methods and metrics should therefore be applied
to additional settings to characterise the overhead-coverage trade-off
under different sampling policies and production conditions.

\section{Related Work}
\label{sec:related-work}

Distributed tracing (DT) provides request-scoped causality across
services and is now a core practice in production
observability~\cite{Parker2020,Shkuro2019}. DT already captures
request propagation through spans, timing, and links, which makes it
the natural scope for energy attribution. It brings end-to-end
causality to complex
systems~\cite{Shkuro2019,Davidson2024}. Workflow-centric
tracing~\cite{Sambasivan2016} and universal context
propagation~\cite{MaceFonseca2018} underpin the design of tracing
tools. This section discusses related work that
leverages DT baggage/context by attaching energy metadata.

By correlating measured power with spans, one can perform fair-share
energy attribution across overlapping work and across services. Using
OpenTelemetry link propagation between causally related spans, it is
possible to preserve
attribution through asynchronous hops and background
work~\cite{Parker2020,otel2024young}.

Reported prototype systems stream per-request estimates through
sidecars and models in FLEET~\cite{FLEET2023},
couple trace graphs with power and network data to recover per-request
totals~\cite{Anand2023OddOne} or unite distributed tracing and
energy/CO\textsubscript{2} by extending the OpenTelemetry Java agent in
RETIT\footnote{\url{https://github.com/RETIT/opentelemetry-javaagent-extension}}~\cite{brunnert2024extending}. These
directions are feasible and useful, and they expose clear trade-offs
between model drift and measurement overhead, and between online
timeliness and post-mortem fidelity.

FLEET~\cite{FLEET2023} augments microservices with sidecars and ML models to estimate
per-request energy and propagate it in-band for policies.
Request energy is transported
primarily in headers rather than embedded as a first-class trace
artefact with formal additivity to measured power.
Our approach differs by favouring \emph{measurement-grounded} attribution, 
via trace timestamps and power correlation with explicit handling of
overlap and idle baselines.

RETIT~\cite{brunnert2024extending} attaches span attributes for CPU time (ms), memory, storage, and
network bytes to every span captured by the standard OpenTelemetry Java
auto-instrumentation. For entry transactions, it also publishes OpenTelemetry
metrics so backends can aggregate Joules per request or endpoint.
Being a Java-only
tool, a practical limitation is span/thread correspondence: values are
valid only if a span starts and ends on the same thread.
In contrast, our approach is language-agnostic, which is common for
microservice architectures.

Unlike technology-dependent (JVM), model-first or
sidecar-centric proposals, we ground attribution in measurements and
make energy a native artefact of traces that existing backends can
query, aggregate, and validate alongside latency and errors, thereby
shifting analysis from an application sidecar concern to a first-class
observability concern. This positions trace-level energy as both an
auditable ledger for post-mortem analysis and a calibration source for
lightweight online estimators that serve policies.

\section{Conclusion}
\label{sec:conclusion}

This paper presented Spanergy, an energy-aware distributed tracing
approach that attributes measured energy consumption to end-user
requests in microservice applications. It adopts a language-agnostic
methodology to correlate per-service power measurements with
OpenTelemetry traces and allocate energy across spans using a
deterministic sweep-line algorithm, enabling request-level energy
estimation that preserves conservation properties and accounts for
both synchronous and asynchronous interactions.
The post-processing adds a controlled incremental cost of
only 15.2\% of the S0 baseline, which
corresponds to roughly 8.7\% of the total S2 energy under the observed
mean ratios, highlighting that Spanergy is lightweight compared to the
cost of enabling tracing itself.

The experimental evaluation demonstrates that Spanergy estimates
per-span and per-request energy, and that it enables endpoint-centric
aggregation into a request-level energy catalogue. This catalogue is
actionable at the API boundary: it ranks endpoints by energy cost and
enables interpretable decompositions of CPU energy consumption per microservice
into demand-driven request types, supporting concrete energy bottleneck
detection, optimisation, and debugging decisions. Spanergy further
enables stable end-user request-level energy estimation (e.g., for the 
\texttt{otel-demo}, \code{POST /api/checkout} at 0.390~J per request) and shows
that energy-aware diagnostics differ from latency-only analysis. While 
latency-critical path (LCP) and energy-critical path (ECP) overlap, a 
non-negligible fraction of energy lies outside the LCP spans, revealing 
energy-dominant work that LCP would miss.

Spanergy enables stakeholders to reason about energy in the same terms
they use for operational and architectural decisions: requests,
endpoints, traces, and journeys. Request-level energy catalogues can
support diverse use cases, from end-user transparency and feature
prioritisation to energy-aware service-level objectives and
carbon-conscious scheduling. In this sense, the catalogue should be read
as a practical showcase of usefulness for decision support in SaaS
governance and operations under the studied conditions.
In our experiments, endpoint energy profiles are stable across
repeated runs for a fixed journey and deployment configuration.
Assessing how well these profiles transfer across other user journeys
and deployment contexts is a natural next step.

A promising direction is to
evaluate sampling policies under explicit constraints, using
conservation and coverage metrics to characterise the trade-off between
telemetry cost and attribution fidelity to determine when robust
energy consumption conclusions can be drawn.
A complementary next step is to broaden telemetry scope beyond CPU power to 
improve attribution coverage across diverse workloads.

\section{Acknowledgements}

This research was produced within the framework of Energy4Climate
Interdisciplinary Center (E4C) of IP Paris.
This research was supported by 3rd Programme
d'Investissements d'Avenir [ANR-18-EUR-0006-02]. This work also
received funding from the France 2030 programme, managed by the French
National Research Agency under grant agreement No. ANR-23-PECL-0003.
The authors would also like to acknowledge the contribution of Henrique de 
Medeiros, who provided helpful comments and expertise on software power meters.



\begin{thebibliography}{10}
\providecommand{\url}[1]{#1}
\csname url@samestyle\endcsname
\providecommand{\newblock}{\relax}
\providecommand{\bibinfo}[2]{#2}
\providecommand{\BIBentrySTDinterwordspacing}{\spaceskip=0pt\relax}
\providecommand{\BIBentryALTinterwordstretchfactor}{4}
\providecommand{\BIBentryALTinterwordspacing}{\spaceskip=\fontdimen2\font plus
\BIBentryALTinterwordstretchfactor\fontdimen3\font minus
  \fontdimen4\font\relax}
\providecommand{\BIBforeignlanguage}[2]{{%
\expandafter\ifx\csname l@#1\endcsname\relax
\typeout{** WARNING: IEEEtran.bst: No hyphenation pattern has been}%
\typeout{** loaded for the language `#1'. Using the pattern for}%
\typeout{** the default language instead.}%
\else
\language=\csname l@#1\endcsname
\fi
#2}}
\providecommand{\BIBdecl}{\relax}
\BIBdecl

\bibitem{Freitag2021ICTClimate}
C.~Freitag, M.~Berners-Lee, K.~Widdicks, B.~Knowles, G.~S. Blair, and
  A.~Friday, ``{The Real Climate and Transformative Impact of ICT},''
  \emph{Patterns}, vol.~2, no.~9, 2021.

\bibitem{KootWijnhoven2021}
M.~Koot and F.~Wijnhoven, ``{Usage Impact on Data Center Electricity Needs: A
  System Dynamic Forecasting Model},'' \emph{Applied Energy}, vol. 291, p.
  116798, 2021.

\bibitem{Andrae2020}
A.~S. Andrae, ``{Hypotheses for Primary Energy Use, Electricity Use and CO2
  Emissions of Global Computing},'' \emph{WSEAS Transactions on Power Systems},
  vol.~15, pp. 50--59, 2020.

\bibitem{Mastelic2014}
T.~Mastelic, A.~Oleksiak, H.~Claussen, I.~Brandic, J.-M. Pierson, and A.~V.
  Vasilakos, ``{Cloud Computing: Survey on Energy Efficiency},'' \emph{ACM
  Computing Surveys}, vol.~47, no.~2, pp. 1--36, 2014.

\bibitem{Orgerie2014}
A.-C. Orgerie, M.~D. Assuncao, and L.~Lefevre, ``{A Survey on Techniques for
  Improving the Energy Efficiency of Large-scale Distributed Systems},''
  \emph{ACM Computing Surveys}, vol.~46, no.~4, pp. 1--31, 2014.

\bibitem{Dayarathna2015}
M.~Dayarathna, Y.~Wen, and R.~Fan, ``{Data Center Energy Consumption Modeling:
  A Survey},'' \emph{IEEE Communications Surveys \& Tutorials}, vol.~18, no.~1,
  pp. 732--794, 2015.

\bibitem{KaurChana2015}
T.~Kaur and I.~Chana, ``{Energy Efficiency Techniques in Cloud Computing: A
  Survey and Taxonomy},'' \emph{ACM Computing Surveys}, vol.~48, no.~2, pp.
  1--46, 2015.

\bibitem{KhanZakarya2021}
A.~A. Khan and M.~Zakarya, ``{Energy, Performance and Cost Efficient Cloud
  Datacentres: A Survey},'' \emph{Computer Science Review}, vol.~40, p. 100390,
  2021.

\bibitem{Buyya2024}
R.~Buyya, S.~Ilager, and P.~Arroba, ``{Energy-efficiency and Sustainability in
  New Generation Cloud Computing},'' \emph{Software: Practice and Experience},
  vol.~54, no.~1, pp. 24--38, 2024.

\bibitem{Hilman2020}
M.~H. Hilman, M.~A. Rodriguez, and R.~Buyya, ``{Multiple Workflows Scheduling
  in Multi-tenant Distributed Systems: A Taxonomy and Future Directions},'' in
  \emph{ACM Computing Surveys}, 2020.

\bibitem{Zhong2022}
Z.~Zhong, M.~Xu, M.~A. Rodriguez, C.~Xu, and R.~Buyya, ``{Machine
  Learning-based Orchestration of Containers: A Taxonomy and Future
  Directions},'' \emph{ACM Computing Surveys}, vol.~54, no. 10s, pp. 1--35,
  2022.

\bibitem{Araujo2024}
G.~Araújo, V.~Barbosa, L.~N. Lima, A.~Sabino, C.~Brito, I.~Fé, P.~Rego,
  E.~Choi, D.~Min, T.~A. Nguyen \emph{et~al.}, ``{Energy Consumption in
  Microservices Architectures: A Systematic Literature Review},'' \emph{IEEE
  Access}, 2024.

\bibitem{Soldani2018}
J.~Soldani, D.~A. Tamburri, and W.-J. Van Den~Heuvel, ``{The Pains and Gains of
  Microservices: A Systematic Grey Literature Review},'' \emph{Journal of
  Systems and Software}, vol. 146, pp. 215--232, 2018.

\bibitem{Haselbock2017}
S.~Haselböck and R.~Weinreich, ``{Decision Guidance Models for Microservice
  Monitoring},'' \emph{ICSAW}, pp. 54--61, 2017.

\bibitem{Parker2020}
A.~Parker, D.~Spoonhower, J.~Mace, and R.~Isaacs, \emph{{Distributed Tracing in
  Practice}}.\hskip 1em plus 0.5em minus 0.4em\relax O'Reilly, 2020.

\bibitem{Davidson2024}
T.~Davidson, E.~Wall, and J.~Mace, ``{A Qualitative Interview Study of
  Distributed Tracing Visualisation},'' \emph{IEEE Transactions on
  Visualization and Computer Graphics}, vol.~30, no.~7, pp. 3828--3840, 2024.

\bibitem{Anand2023OddOne}
V.~Anand, Z.~Xie, M.~Stolet, R.~De~Viti, T.~Davidson, R.~Karimipour,
  S.~Alzayat, and J.~Mace, ``{The Odd One Out: Energy Is Not Like Other
  Metrics},'' \emph{ACM SIGENERGY Energy Informatics Review}, vol.~3, no.~3,
  pp. 71--77, 2023.

\bibitem{otel2024young}
T.~Young and A.~Parker, \emph{{Learning OpenTelemetry}}.\hskip 1em plus 0.5em
  minus 0.4em\relax O'Reilly Media, Inc., 2024.

\bibitem{otel}
OpenTelemetry, \url{https://opentelemetry.io/docs/}, accessed December 2025.

\bibitem{MysteryMachine2014}
M.~Chow, D.~Meisner, J.~Flinn, D.~Peek, and T.~F. Wenisch, ``{The Mystery
  Machine: End-to-end Performance Analysis of Large-scale Internet Services},''
  in \emph{11th USENIX Symposium on Operating Systems Design and Implementation
  (OSDI)}, Oct. 2014, pp. 217--231.

\bibitem{CRISP2022}
Z.~Zhang, M.~K. Ramanathan, P.~Raj, A.~Parwal, T.~Sherwood, and M.~Chabbi,
  ``{CRISP: Critical Path Analysis of Large-Scale Microservice
  Architectures},'' in \emph{2022 USENIX Annual Technical Conf. (ATC)}, Jul.
  2022, pp. 655--672.

\bibitem{Legler2025}
J.~Legler, S.~Werner, M.~C. Borges, and S.~Tai, ``{Service-Level Energy
  Modeling and Experimentation for Cloud-Native Microservices},'' in \emph{23rd
  Int. Conf. on Service-Oriented Computing (ICSOC)}, ser. LNCS, vol.
  16320.\hskip 1em plus 0.5em minus 0.4em\relax Shenzhen, China: Springer, Dec.
  2025.

\bibitem{Wohlin2012Experimentation}
C.~Wohlin, P.~Runeson, M.~H{\"o}st, M.~C. Ohlsson, B.~Regnell, and
  A.~Wessl{\'e}n, \emph{{Experimentation in Software Engineering}}.\hskip 1em
  plus 0.5em minus 0.4em\relax Springer, 2012.

\bibitem{ArcuriBriand2014Hitchhikers}
A.~Arcuri and L.~Briand, ``A hitchhiker's guide to statistical tests for
  assessing randomized algorithms in software engineering,'' \emph{Software
  Testing, Verification and Reliability}, vol.~24, no.~3, pp. 219--250, 2014.

\bibitem{KitchenhamMadeyski2017Robust}
B.~Kitchenham, L.~Madeyski, D.~Budgen, J.~Keung, P.~Brereton, S.~Charters,
  S.~Gibbs \emph{et~al.}, ``{Robust Statistical Methods for Empirical Software
  Engineering},'' \emph{Empirical Software Engineering}, vol.~22, no.~2, pp.
  576--630, 2017.

\bibitem{Dolado2014Equivalence}
J.~J. Dolado, M.~C. Otero, and M.~Harman, ``Equivalence hypothesis testing in
  experimental software engineering,'' \emph{Software Quality Journal},
  vol.~22, no.~2, pp. 215--238, 2014.

\bibitem{ShapiroWilk}
S.~S. Shapiro and M.~Wilk, ``{An Analysis of Variance Test for Normality},''
  \emph{Biometrika}, vol.~52, no. 3-4, pp. 591--–611, 1965.

\bibitem{Welch}
B.~L. Welch, ``{The generalization of "Student's" problem when several
  different population variances are involved},'' \emph{Biometrika}, vol.~34,
  no. 1-2, pp. 28--–35, 1947.

\bibitem{Shkuro2019}
\BIBentryALTinterwordspacing
Y.~Shkuro, \emph{{Mastering Distributed Tracing: Analyzing Performance in
  Microservices and Complex Systems}}.\hskip 1em plus 0.5em minus 0.4em\relax
  Packt Publishing, 2019. [Online]. Available:
  \url{https://www.oreilly.com/library/view/mastering-distributed-tracing/9781788628464/}
\BIBentrySTDinterwordspacing

\bibitem{Sambasivan2016}
R.~R. Sambasivan, I.~Shafer, J.~Mace, B.~H. Sigelman, R.~Fonseca, and G.~R.
  Ganger, ``{Principled Workflow-centric Tracing of Distributed Systems},'' in
  \emph{SoCC}, 2016, pp. 401--414.

\bibitem{MaceFonseca2018}
J.~Mace and R.~Fonseca, ``{Universal Context Propagation for Distributed System
  Instrumentation},'' in \emph{EuroSys}, 2018.

\bibitem{FLEET2023}
\BIBentryALTinterwordspacing
C.~Meadows, ``{FLEET: Fine‐grained, Lightweight Energy Estimate Tracing for
  Microservice Architectures},'' Master's thesis, The George Washington
  University, USA, 2023. [Online]. Available:
  \url{https://scholarspace.library.gwu.edu/concern/gw_etds/5t34sk50c}
\BIBentrySTDinterwordspacing

\bibitem{brunnert2024extending}
A.~Brunnert and F.~Gutzy, ``{Extending the OpenTelemetry Java
  Auto-Instrumentation Agent to Publish Green Software Metrics},'' in
  \emph{Softwaretechnik-Trends, 44(4)}, 2024.

\end{thebibliography}
\end{document}